\documentclass{article}

\usepackage{arxiv}
\usepackage[utf8]{inputenc} 
\usepackage[T1]{fontenc} 
\usepackage[english]{babel}
\usepackage{booktabs} 
\usepackage{amsfonts} 
\usepackage{amsmath,amssymb,latexsym}
\usepackage{microtype} 
\usepackage{color}
\usepackage{fullpage}
\usepackage{graphicx}
\usepackage{morefloats}
\usepackage{lscape}
\usepackage{fancyhdr}
\usepackage{rotating}
\usepackage{bookmark}
\usepackage{xspace}

\newcommand\nifsx{${}^{56}$Ni\xspace}
\newcommand\cofsx{${}^{56}$Co\xspace}
\newcommand\fefsx{${}^{56}$Fe\xspace}

\newcommand{\code}[1]{\texttt{#1}}

\title{Analytical Model of Time-Dependent Ionization in the Envelopes of Type II Supernovae at the Photospheric Phase}

\date{2019 year}

\author
{
  M.Sh.~Potashov \\
  NRC ``Kurchatov Institute'' - ITEP, ul. Bolshaya Cheremushkinskaya 25, Moscow 117218, Russia\\
  Novosibirsk State University, Pirogova 1, Novosibirsk 630090, Russia\\
  \texttt{marat.potashov@gmail.com} \\
  \And
  S.I.~Blinnikov \\
  NRC ``Kurchatov Institute'' - ITEP, ul. Bolshaya Cheremushkinskaya 25, Moscow 117218, Russia\\
  Space Research Institute, Russian Academy of Sciences, Profsoyuznaya ul. 84/32, Moscow, 117997 Russia\\
  Institute of Physics and Mathematics of the Universe, University of Tokyo, Kashiwa, Japan\\
  \texttt{sblinnikov@gmail.com} \\
}

\begin{document}

\maketitle

\begin{abstract}
  We investigate a simplified kinetic system of the hydrogen atom
    (two levels plus continuum) under conditions of a type~IIP supernova
    at the plateau phase that realistically describes the basic properties
    of the complete system.
  We have found the Lyapunov function for the reduced system
    using which we have analytically obtained the ionization freeze-out
    effect on long time scales.
  Since the system completely recombines in the equilibrium
    approximation on long time scales,
    which does not occur in reality,
    this result confirms the necessity of allowance
    for the time-dependent effect in the kinetics during
    the photospheric phase in a supernova explosion.
\end{abstract}



\section*{INTRODUCTION AND FORMULATION OF THE PROBLEM}
  \label{sec:introduction}

New data, the photometric distances to objects with known redshifts,
  are required to investigate the present-day structure of the Universe.
Among the great variety of distance measurement techniques
  there are methods that do not rely on the cosmological distance ladder,
  for example, the expanding photosphere method
  (EPM) \cite{KirshnerKwan1974},
  the spectral-fitting expanding atmosphere method
  (SEAM) \cite{BaronNugentBranchEtal2004},
  or the dense shell method (DSM)
  \cite{BlinnikovPotashovBaklanovEtal2012,
    PotashovBlinnikovBaklanovEtal2013,
    BaklanovBlinnikovPotashovEtal2013a}
  that use type IIP and IIn supernovae (SNe) as objects.
Using such a method as the SEAM requires the construction
  of a complete physical model for a type~II SN
  that reproduces in detail its spectrum.

The importance of direct cosmological distance measurement methods
  is particularly topical in light of the problem of an uncertainty
  in measuring the Hubble parameter (Hubble tension)
  \cite{RiessCasertanoYuanEtal2018,
    MortsellDhawan2018,
    EzquiagaZumalacarregui2018}.

To completely model the physical processes occurring in a SN,
  it is necessary to simultaneously take into account
  the envelope expansion hydrodynamics,
  the matter--radiation field interaction,
  the radiative transfer in lines and continuum,
  and the kinetics of level populations in the atoms of a multiply charged plasma.
This gives a system of integro-differential equations
  of radiation hydrodynamics that cannot yet be completely solved numerically
  even in the one-dimensional case.
One has to resort to unavoidable simplifications in this complete system.
One of such simplifications is the steady-state approximation
  of the kinetic system of level populations,
  when the system is assumed to be in statistical equilibrium.

The time-dependent hydrogen ionization effect in the envelopes
  of type~II SNe at the photospheric phase was used
  by Kirshner and Kwan \cite{KirshnerKwan1975}
  to explain the high H$\alpha$ intensity in the spectra of SN~1970G
  and by Chugai \cite{Chugai1991}
  to explain the high degree of hydrogen excitation
  in the outer atmospheric layers ($v>7000$ km\,s$^{-1}$)
  of SN~1987A in the first 40 days after its explosion.

Utrobin and Chugai \cite{UtrobinChugai2002}
  found a strong time-dependent effect
  in the ionization kinetics and hydrogen lines in type~IIP SNe
  during the photospheric phase.
In their next paper Utrobin and Chugai \cite{UtrobinChugai2005}
  also took into account the time-dependent effect in the energy equation.
An important consequence of these papers was the conclusion
  that including the time-dependent ionization allowed the spectra
  of peculiar SN~1987A with a stronger H$\alpha$ line to be obtained,
  which could not be done previously without
  mixing radioactive \nifsx into the outer high-velocity layers
  in the steady-state approximation.
In the next paper \cite{Utrobin2007} the importance of this effect
  was also shown for normal SN~1999em.

The conclusions reached by Utrobin and Chugai
  were confirmed by Dessart and Hillier using the \code{CMFGEN} software package.
In Dessart et al. \cite{DessartBlondinBrownEtal2008}
  the applied approach was still the steady-state one
  and it was implemented in the \code{CMFGEN} package.
Modeling revealed a problem: the H$\alpha$ line in hydrogen-rich envelopes
  was weaker than that observed at the recombination epoch.
In particular, the model did not reproduce the line for times
  later than four days for SN~1987A and later than 20~days for SN~1999em.
Next, Dessart and Hillier improved the code by including the time dependence
  in the kinetic system and the energy equation
  \cite{DessartHillier2007}
  and then in the radiative transfer
  \cite{DessartHillier2010a, HillierDessart2012}.
This allowed the H$\alpha$ line to be strengthened in the resulting spectrum,
  which led to better agreement with observations.

On the other hand, based on their computations with the
  \code{PHOENIX} software package,
  De et al. \cite{DeBaronHauschildt2010a}
  found the time-dependent kinetics
  to be important only in the first days after SN explosion.
Moreover, they argue that the role of the time-dependent effect
  is not very strong even in these first days by illustrating this with
  the models of SN~1987A and SN~1999em as an example.
Using the open \code{TARDIS} code and without negating
  the importance of the time-dependent effect in the kinetics,
  Vogl et al. \cite{VoglSimNoebauerEtal2019}
  nevertheless neglect it when modeling the spectra
  of SN~1999em and obtain good agreement of them with the observed ones.
The overwhelming majority of Monte Carlo simulation codes also neglect
  the time-dependent effect in the kinetics.
Thus, the conclusions of various research groups disagree
  and the importance of this effect is still called into question.

Using the \code{STELLA} and \code{LEVELS} codes,
  Potashov et al. \cite{PotashovBlinnikovUtrobin2017a}
  showed the importance of the time-dependent kinetics
  in the purely hydrogen case for SN~1999em as an example.
The influence of metal admixtures on the intensity of the effect was also investigated:
  an increase in the concentration of metals in the envelope
  led to a weakening of the time-dependent effect in the kinetics.

In this paper we will attempt to answer the question of whether
  the time-dependent ionization effect is important or not,
  at least on long time scales, within a simple analytical model.


\section*{MODELING}
  \label{sec:model}

Let us describe the construction of a simple analytical model
  for the behavior of multiply charged plasma electron populations in a SN envelope.
  We consider a purely hydrogen envelope,
  where the hydrogen atom is represented by a ``two levels + continuum'' system.
We assume an $l$-equilibrium for the second atomic level.
This means that the populations of sublevels including the fine structure
  of $2s$ and $2p$ are proportional to their statistical weights.
Thus, the second level is considered as a single superlevel
   \cite{HubenyLanz1995}.

The characteristic stages of the behavior of the light curve
  for a typical type~IIP SN can be written as follows \cite{Utrobin2007}:
\begin{itemize}
  \item shock breakout;
  \item the adiabatic cooling phase;
  \item the photospheric phase (a cooling and recombination wave is formed);
  \item the radiative diffusion cooling
    (the radiative diffusion time is less than the characteristic
    envelope expansion time);
  \item the beginning of thermal energy exhaustion;
  \item the end of thermal energy exhaustion (plateau tail phase);
  \item the nonthermal emission due to the decays
    \nifsx $\rightarrow$ \cofsx $\rightarrow$ \fefsx (radioactive tail).
\end{itemize}

To study the time-dependent hydrogen ionization
  in the envelope plasma, we will consider the behavior of the system
  only at the photospheric phase.
For typical SN~1999em
  \cite{BaklanovBlinnikovPavlyuk2005, Utrobin2007}
  this phase lasts from $t_0 \sim 20$ to $t_1 \sim 100$ days.
A cooling and hydrogen recombination wave is formed as the envelope expands.
The bolometric luminosity at the outer boundary of this wave
  is equal to the luminosity of the entire star.
The photosphere is also located at the same boundary.
An important fact is that in this period the photospheric radius $R_{ph}$,
the radiation temperature $T_c$,
and the matter temperature $T_e$ remain almost constant.
Consequently, the total luminosity of the star does not change
  with time and the light curve reaches a plateau.

Our subsequent description suggests that $t \geqslant t_0$.
Our modeling with the \code{STELLA} code shows that
  the transition to a homologous (with a high accuracy)
  expansion of the SN~1999em envelope ends approximately
  by day~15 after explosion \cite{BaklanovBlinnikovPavlyuk2005},
  i.e., earlier than the beginning of the
  photospheric phase $t_0$.
We assume an isotropic spherically symmetric expansion.
In our initial analysis we also disregard
  the collisional excitation and ionization processes.

Let us select some small region of the envelope above the photosphere.
The continuity equation in Eulerian coordinates for the matter in this region is
\begin{equation}
  \frac{\partial\rho}{\partial t} = -\bigtriangledown(\rho v),
\label{eq:continuity:eulerian}
\end{equation}
where $\rho$ is the density of the envelope expanding with a velocity $v$.
In the Lagrangian formalism in the comoving frame we obtain
\begin{equation}
  \frac{D\rho}{D t} = -\rho(\bigtriangledown \cdot v)
\label{eq:continuity:lagrangean}
\end{equation}
In the period of a free homologous expansion Eq. (\ref{eq:continuity:lagrangean})
  is simplified to
\begin{equation}
  \frac{D\rho}{D t} + \frac{3\rho}{t} = 0.
\label{eq:continuity:lagrangean:homogeneity}
\end{equation}
The rate of transitions to any discrete bound or free level $i$
  of a hydrogen atom or ion can then be written as
\begin{equation}
  \frac{Dn_{i}}{D t} + \frac{3n_{i}}{t} = K_{i}(t),
\label{eq:kinetics}
\end{equation}
where $n_{i}$ is the population of level $i$ in the atom or ion.
In turn, neglecting the stimulated emission processes,
  we define the function $K_i(t)$ as
\begin{equation}
  K_{1}(t) = (N(t) - n_1 - n_e)\;(Q + A_{21}) + n_1B_{12}J_{12}(t)
  \label{eq:kinetics:K:H0}
\end{equation}
\begin{equation}
  K_{e}(t) = (N(t) - n_1 - n_e)\;P_{2c}(t) - n_e^2\;R_{c2}(t)
  \; .
\label{eq:kinetics:K:H1}
\end{equation}
Here,
\begin{equation}
  N(t) = {N_0}\frac{t_0^3}{t^3}
\end{equation}
  is the hydrogen number density;
  $Q$ is the two-photon $2s \to 1s$ decay rate;
  The reverse $1s \to 2s$ transition (two-photon absorption) rate
    is much lower than the $2s \to 1s$ rate
    and we disregard this process \cite{PotashovBlinnikovUtrobin2017a};
  $A_{21}$ and $B_{12}$ are the Einstein coefficients
    for the spontaneous emission and photoexcitation
    of the $1 \leftrightarrow 2$ transition;
  $J_{12}(t)$ is the intensity of radiation in the $2 \to 1$ transition
    averaged over the line profile;
  $P_{2c}(t)$ is the total photoionization coefficient from the second level;
  $R_{c2}(t)$ is the total radiative recombination coefficient to the second level.
  
In our model we use the fact that the bound--free processes make a major
  contribution to the opacity in the Lyman continuum
  frequency band $\nu \geqslant \nu_{LyC}$
  \cite{PotashovBlinnikovUtrobin2017a}.
We neglect the relatively small contributions of the bound-bound processes
  in lines (the so-called expansion opacity)
  and free--free processes to the emission and absorption coefficients.
The absorption in this band is caused mainly by neutral hydrogen
  and the optical depth is very large.
Therefore, there is virtually no photospheric radiation here
  and the radiation field is determined for the regions above the photosphere
  by diffusive radiation.
In this case, it can be shown that the photoionization rate from
  the ground hydrogen level and the recombination rate
  to the ground level closely coincide
  (even if there is not only hydrogen in the envelope).
Thus, the first hydrogen level is in detailed balance
  with the continuum and the corresponding processes
  do not enter into the system of equations
  (\ref{eq:kinetics:K:H0}), (\ref{eq:kinetics:K:H1}).

It should be noted that in the Lyman continuum frequency band
  the intensity of continuum diffusive radiation $J_c(\nu)$
  coincides with the equilibrium one $B_{\nu}(T_e)$
  only in a purely hydrogen envelope,
  suggesting a matter--radiation equilibrium.
In the general case, with admixtures, $J_c(\nu) \neq B_{\nu}(T_e)$.

Let us write out the standard formulas of Sobolev’s approximation
  \cite{Sobolev1960book, Castor1970},
  but in a simplified form, using the condition that the population
  of the second level is relatively small,
  $N(t) - n_1 - n_e \ll n_1$.

The Sobolev optical depth is
\begin{equation}
  \tau(t) \sim \frac{c^3}{8\pi}\frac{A_{21}}{\nu_{L\alpha}^3}\frac{g_2}{g_1}n_1t,
\end{equation}
  the intensity of radiation averaged over the profile and the angles is
\begin{equation}
  J_{12}(t) = (1 - \beta(t)) \; S(t) + \beta(t) \, J_c(\nu_{L\alpha}, t),
\label{eq:jlu}
\end{equation}
  where $J_c(\nu_{L\alpha})$ is the intensity of continuum radiation
  at the L$\alpha$ frequency.

We assume that $\tau(t) \gg 1$.
The local escape probability of an L$\alpha$ photon
  without scattering integrated over the directions
  and over the line frequencies is then
\begin{equation}
  \beta(t) = \frac{1-e^{-\tau(t)}}{\tau(t)}\sim\frac{1}{\tau(t)}.
\label{eq:beta}
\end{equation}
The source function is
\begin{equation}
  S_{12}(t) = \frac{2 h \nu_{L\alpha}^3}{c^2},
    \left( \frac{g_2 n_1}{g_1 n_2} \right)
\label{eq:slu}
\end{equation}
  all the remaining notation is the standard one.

For an optically thick (in the L$\alpha$ line)
  SN envelope the estimate (\ref{eq:beta}) breaks down.
The absorption of photons in continuum should be taken into account
  \cite{HummerRybicki1985, Chugai1987}.
However, applying these corrections will not change
  qualitatively the final result of this paper.

Combining (\ref{eq:kinetics})--(\ref{eq:kinetics:K:H1})
  and (\ref{eq:jlu})--(\ref{eq:slu}), we obtain the system
\begin{equation}
  \dot{n_{1}}  =
    (N(t) - n_1 - n_e)\;(Q + A_{21}\beta(t))
    - n_1B_{12}\beta(t) J_c(\nu_{L\alpha}, t) - \frac{3n_1}{t}
\end{equation}
\begin{equation}
  \dot{n_{e}} = (N(t) - n_1 - n_e)\;P_{2c}(t) - n_e^2\;R_{c2}(t) - \frac{3n_e}{t}
\end{equation}
According to Mihalas \cite{HubenyMihalas2014book}
  the photoionization rate is the integral
\begin{equation}
    P_{2c}(t) = 4\pi\int_{\nu_2}^{\infty} \alpha_{12}(\nu)\frac{J_c(\nu, t)}{h\nu}d\nu.
\end{equation}
while the photorecombination rate for a purely hydrogen plasma
  in the case where the stimulated emission is neglected will appear as
\begin{equation}
  \begin{split}
    R_{c2}
    = \;&4\pi\Phi_{\mathrm{Saha}}(T_e)\;\int_{\nu_2}^{\infty} \alpha_{12}(\nu)\frac{1}{h\nu}\frac{2h\nu^3}{c^2}e^{-\frac{h\nu}{kT_e}}d\nu
    \sim \\
    & \Phi_{\mathrm{Saha}}(T_e)\;g_{II}(1,\nu_{L\alpha})\frac{\pi}{c^2}E_1\left(\frac{h\nu_2}{kT_e}\right)
  \end{split}
\end{equation}
Here
  $\Phi_{\mathrm{Saha}}(T_e)$ is the Saha factor;
  $g_{II}(1,\nu_{L\alpha})$ is the Gaunt factor for the bound--free transition $1 \leftrightarrow 2$,
  and $E_1$ is a modified exponential integral.
It can be noted that the constancy of $R_{c2}$
  in time follows from the constancy of $T_e$.

Let us now introduce dimensionless variables:
\begin{equation}
  u_1 = \frac{n_1}{N(t)} = \frac{n_1}{N_0}\frac{t^3}{t_0^3}, \qquad
  u_e = \frac{n_e}{N(t)} = \frac{n_e}{N_0}\frac{t^3}{t_0^3},
  \nonumber
\end{equation}
  which are the level populations normalized
  to the total current number density.

Rewriting the system in them, we obtain
\begin{equation}
  \dot{u_{1}}  = (1-u_1-u_e)\;\left(Q + \frac{A}{u_1}\left(\frac{t}{t_0}\right)^2\right) -
  \tilde B J_c(\nu_{L\alpha}, t) \left(\frac{t}{t_0}\right)^2
\label{eq:kinetics:u1:td1}
\end{equation}
\begin{equation}
  \dot{u_{e}}  = (1-u_1-u_e)\;P_{2c}(t)-u_e^2R\left(\frac{t_0}{t}\right)^3
\label{eq:kinetics:ue:td1}
\end{equation}
New notation is also introduced here:
\begin{equation}
  A = \frac{8 \pi \nu_{L\alpha}^3}{c^3} \frac{g_1}{g_2} \cdot \frac{1}{N_0 t_0}, \quad
  \tilde B = \frac{4\pi}{h\,c} \cdot \frac{1}{N_0 t_0}, \quad
  R = N_0 R_{c2}
  \nonumber
\end{equation}
Investigating the behavior of $J_c(\nu_{L\alpha}, t)$ and $P_{2c}(t)$
  with time is fundamentally important for a further simplification
  of system (\ref{eq:kinetics:u1:td1}), (\ref{eq:kinetics:ue:td1}).
In the optically thin case, we can write $J_c(t) = W(t) B(T_c)$.
If, in addition, we assume that the region under consideration
  is sufficiently far from the photosphere,
  then the dilution factor will change with time as
\begin{equation}
  W(t) \sim \frac{1}{4}\left(\frac{R_{ph}}{Vt}\right)^2
\end{equation}
The continuum intensity $J_c(\nu_{L\alpha}, t)$
  and the photoionization rate $P_{2c}(t)$ will then drop as $\propto 1/t^2$.

In the observed SN the medium at the frequencies under consideration
  in continuum is optically thick.
A large amount of metal admixtures changes the behavior
  of the radiation intensity in the hard band by reducing it.
However, numerical simulations (for example, using the \code{STELLA} code)
  show a power-law time dependence of the intensity
  and the photoionization rate even in this case.
Specifically,
\begin{equation}
  J_c(\nu_{L\alpha}, t) = J_c(\nu_{L\alpha}, t_0) \left(\frac{t_0}{t}\right)^{s_1}
  \label{eq:jc}
\end{equation}
\begin{equation}
  P_{2c}(t) = P_{2c}(t_0) \left(\frac{t_0}{t}\right)^{s_2} = P \left(\frac{t_0}{t}\right)^{s_2}
  \label{eq:p2c}
\end{equation}
The exponents $s_1$ and $s_2$ depend on the distance
  from the photosphere, but they are always greater than 2.
Thus, generally, we restrict the domain of definition
  of the powers as $s_1 \geqslant 2$ and $s_2 \geqslant 2$.

System (\ref{eq:kinetics:u1:td1}), (\ref{eq:kinetics:ue:td1}),
  given (\ref{eq:jc}), (\ref{eq:p2c}), will appear as
\begin{equation}
  \dot{u_{1}}^{td}  = (1-u_1^{td}-u_e^{td})\;\left(Q + \frac{A}{u_1^{td}}\left(\frac{t}{t_0}\right)^2\right) -
  B \left(\frac{t_0}{t}\right)^{s_1-2}
\label{eq:kinetics:u1:td2}
\end{equation}
\begin{equation}
  \dot{u_{e}}^{td}  = (1-u_1^{td}-u_e^{td})\;P\left(\frac{t_0}{t}\right)^{s_2} - (u_e^{td})^2R\left(\frac{t_0}{t}\right)^3,
\label{eq:kinetics:ue:td2}
\end{equation}
  where $B = \tilde B J_c(\nu_{L\alpha}, t_0)$.

The equilibrium populations in the same approximation
  can be found by solving the following system
  of algebraic equations:
\begin{equation}
  (1-u_1^{ss}-u_e^{ss})\;\left(Q + \frac{A}{u_1^{ss}}\left(\frac{t}{t_0}\right)^2\right) -
    B \left(\frac{t_0}{t}\right)^{s_1-2} = 0
\label{eq:kinetics:u1:ss}
\end{equation}
\begin{equation}
  (1-u_1^{ss}-u_e^{ss})\;P\left(\frac{t_0}{t}\right)^{s_2} - (u_e^{ss})^2R\left(\frac{t_0}{t}\right)^3 = 0.
\label{eq:kinetics:ue:ss}
\end{equation}

Thus, an answer to the question about the importance
  of allowance for the time-dependent effect in the kinetics
  should be sought by comparing the solutions
  of systems (\ref{eq:kinetics:u1:td2}, \ref{eq:kinetics:ue:td2})
  and (\ref{eq:kinetics:u1:ss}, \ref{eq:kinetics:ue:ss})
  or (equivalently)
  $u_1^{td}$, $u_e^{td}$ and $u_1^{ss}$, $u_e^{ss}$,
  respectively.

Below we show that any physically reasonable
  bounded solution of (\ref{eq:kinetics:u1:td2}, \ref{eq:kinetics:ue:td2})
  is Lyapunov stable ``in the small'',
  i.e., stability is guaranteed at sufficiently small deviations.

Let we know one of the bounded solutions
  $0 < \tilde u_1 \leqslant 1$ and $0 \leqslant \tilde u_e \leqslant 1$
  (\emph{unperturbed motion})
  of the non-autonomous nonlinear differential system
  (\ref{eq:kinetics:u1:td2}, \ref{eq:kinetics:ue:td2}).
Let $x = \tilde u_1 - u_1$ and $y = \tilde u_e - u_e$,
  i.e., $x$ and $y$ are the deviations of the solutions
  $u_1$, $u_e$ from $\tilde u_1$ and $\tilde u_e$, respectively.

For $x$ and $y$ we then obtain a reduced system
  of differential equations
  (it is called the Lyapunov system of equations of \emph{perturbed motion})
  \cite{Demidovich1967book, Khalil2002book}:
\begin{equation}
  \dot{x}  = -(x+y)\;Q - \frac{(1-\tilde u_e)x+\tilde u_1y}{\tilde u_1(\tilde u_1+x)}A\left(\frac{t}{t_0}\right)^2
\label{eq:kinetics:x}
\end{equation}
\begin{equation}
  \dot{y}  = - (x+y)\;P\left(\frac{t_0}{t}\right)^{s_2}-y\;(2\tilde u_e+y)R\left(\frac{t_0}{t}\right)^3.
\label{eq:kinetics:y}
\end{equation}

It is important to note that the trivial solution
  $x = 0$, $y = 0$ is an equilibrium.
Thus, investigating the Lyapunov stability of the solution
  $\tilde u_1$, $\tilde u_e$ is reduced to investigating
  the Lyapunov stability of the trivial solution (equilibrium position)
  $x = 0$, $y = 0$.

Let us next consider a scalar Lyapunov function of the following form:
\begin{equation}
  V(t, x, y) = x^2+2 x y + y^2 \left(2 + \frac{B^2}{A} \frac{1}{\tilde u_1} \frac{t_0^2}{t}\right).
\label{eq:v}
\end{equation}
Obviously, (\ref{eq:v}) is positive definite for all instants of time.
In view of the linearized system (\ref{eq:kinetics:x}, \ref{eq:kinetics:y}),
  its time derivative can be written as
\begin{equation}
  \dot{V}(t, x, y) = \frac{\partial V}{\partial t} + \frac{\partial V}{\partial x} \dot{x} + \frac{\partial V}{\partial y} \dot{y},
\label{eq:dv}
\end{equation}
  where $\dot{x}$ and $\dot{y}$
  are (\ref{eq:kinetics:x}) and (\ref{eq:kinetics:y}), respectively.
It is a negative-definite quadratic form according
  to Sylvester’s criterion, because its corner minors when $t \rightarrow \infty$ are
\begin{equation}
  \Delta_1 \propto - \frac{2A}{\tilde u_1}\left(\frac{t}{t_0}\right)^2 < 0, \qquad
  \Delta_2 \propto \left(\frac{B}{\tilde u_1}\right)^2 > 0.
  \nonumber
\end{equation}
The derivative (\ref{eq:dv1}) itself is negative in sign on long time scales:
\begin{equation}
  \dot{V}(t, x, y) = - 2 A\frac{(x+y)^2}{\tilde u_1} \left(\frac{t}{t_0}\right)^2 + \mathcal{O}(t).
\label{eq:dv1}
\end{equation}
Hence, according to Lyapunov’s first stability theorem
  \cite{Demidovich1967book, Khalil2002book},
  the trivial solution of system (\ref{eq:kinetics:x}, \ref{eq:kinetics:y})
  is Lyapunov stable ``in the small''.
It should be noted that for the system under consideration the stability problem
  can be solved by this method on the semiaxis $t > \tilde t$
  with a sufficiently distant boundary $\tilde t \geqslant t_0$.
Stability on the pre-specified semiaxis $t > t_0$
  is obtained by taking into account the known results
  on continuity in parameter
  \cite{Petrovsky1984book}
  for the solution on the finite interval $t_0 \leqslant t \leqslant \tilde t$.

It can be shown that system
  (\ref{eq:kinetics:x}) and (\ref{eq:kinetics:y})
  is dissipative using Yoshizawa’s theorem
  \cite{Demidovich1967book, KuncevichLychak1977book},
  i.e., the system is also stable ``in the large''.
Consequently, all solutions of system
  (\ref{eq:kinetics:u1:td2}), (\ref{eq:kinetics:ue:td2})
  with physically reasonable initial conditions are bounded always.

Due to the boundedness in $u_{e}^{td}$, it follows from (\ref{eq:kinetics:ue:td2}) that
\begin{equation}
    \lim_{t\to\infty} \dot{u_{e}}^{td} = 0.
    \nonumber
\end{equation}
In turn, by solving (\ref{eq:kinetics:u1:ss}), (\ref{eq:kinetics:ue:ss}),
  it can be shown that
\begin{equation}
  \lim_{t\to\infty} \frac{\partial\;\ln \;u_e^{ss}}{\partial\;\ln\;t} = -\frac{s_1 + s_2 - 3}{2}.
  \nonumber
\end{equation}

Hence it follows that on long time scales the true
  relative electron number density reaches a constant,
  $u_e^{td} \sim c_1$, while the equilibrium relative
  electron number density approaches zero as $u_e^{ss} \sim t^{-(s_1 + s_2 - 3)/2}$.

It can be seen that in the unsteady-state case the envelope expands with
  a higher degree of ionization than in the steady-state approximation.
This should be taken into account when modeling the SN kinetics.


\section*{CONCLUSIONS}
  \label{sec:conclusions}

Such a phenomenon is also observed in atmospheric explosions
  \cite{Raizer1959, ZeldovichRajzer2008book}
  and during the ``protraction'' of primordial plasma recombination
  in the early Universe under cosmological conditions
  \cite{ZeldovichKurtSunyaev1968, Peebles1968}.
The number density of free electrons is commonly said to experience ``freeze-out.''
However, in contrast to the classical freeze-out in atmospheric explosions,
  the time-dependent effect in SNe remains even
  when the temperatures of both matter and radiation are constant.

In this paper we considered the breakdown of the equilibrium steady-state
    approximation itself in the kinetics.
The magnitude of this breakdown and its evolution with time will be
  described in subsequent publications.


\section*{ACKNOWLEDGMENTS}
  \label{sec:acknowledgments}

We are grateful to N.N.~Shakhvorostova,
  V.P.~Utrobin, and A.V.~Yudin for the stimulating discussions.


\section*{FUNDING}
  \label{sec:funding}

The work of M.Sh.~Potashov was supported in part by
  RFBR grant no.~19--02--00567,
  while the work of S.I.~Blinnikov on numerical SN models
  was supported by RSF grant no.~18--12--00522.



\vspace{0.5cm}\footnotesize\noindent
\textit{Translated by V. Astakhov}.

\end{document}